\documentclass[a4paper]{jpconf}
\usepackage{graphicx}

\usepackage{graphicx,color}
\usepackage{graphics}
\usepackage{rotating}
\usepackage{amssymb}

\usepackage{color}

\usepackage{epsfig}

\def\be{\begin{equation}}
\def\ee{\end{equation}}
\def\ba{\begin{eqnarray}}
\def\ea{\end{eqnarray}}
\def\bea{\begin{eqnarray}}
\def\eea{\end{eqnarray}}
\def\bes{\begin{subequations}}
\def\ees{\end{subequations}}
\def\bear{\begin{array}}
\def\eear{\end{array}}

\def \ms {{\overline{\mbox{MS}}}}

\usepackage{epstopdf}

\newcommand{\A}{{\mathcal{A}}}

\newcommand{\MSbar}{\overline{\rm MS}}

\begin{document}
\title{Bjorken sum rule in QCD with analytic coupling
}

\author{C  Ayala$^1$, G  Cveti\v{c}$^1$, A  V~Kotikov$^{2}$
and B G~Shaikhatdenov$^2$}

\address{$^1$ Department of Physics, Universidad T{\'e}cnica Federico
Santa Mar{\'\i}a
Valpara{\'\i}so, Chile}
\address{$^2$ Joint Institute for Nuclear Research, $141980$ Dubna, Russia}

\ead{kotikov@theor.jinr.ru}




\begin{abstract}
We present details of study of the Bjorken polarized sum rule carried out recently in~\cite{ACKS} within the range of energies where the data were collected by JLAB collaboration, $0.05 \ {\rm GeV}^2 <Q^2< 3 \ {\rm GeV}^2$. Three approaches to QCD with analytic (holomorphic) coupling are considered: Analytic Perturbation Theory (APT), Two-delta analytic QCD (2$\delta$anQCD), and Three-delta lattice-motivated analytic QCD in the three-loop and four-loop MiniMOM schemes (3l3$\delta$anQCD, 4l3$\delta$anQCD).
The new frameworks (2$\delta$ and 3$\delta$) with respective couplings give results which agree well with the experimental data for $0.5 \ {\rm GeV}^2<Q^2<3 \ {\rm GeV}^2$ already when only one higher-twist term is taken into account.
\end{abstract}

\section{Introduction}

\label{sec:intr}
The well-known Bjorken polarized sum rule (BSR) $\Gamma_1^{p-n}(Q^2)$ \cite{BjorkenSR} is chosen for testing the nonperturbative behavior in QCD. The experimental data on this quantity published by Jefferson Lab (JLAB) cover the low-$Q^2$ the range $0.05 \ {\rm GeV}^2 < Q^2 < 3 \ {\rm GeV}^2$ \cite{data2,data2B,data2N},
and SLAC data \cite{data3} have $1 \ {\rm GeV}^2 < Q^2$. The theoretical perturbation expansion of the leading-twist (LT) contribution to BSR is known to ${\rm N}^3{\rm LO}$ ($\sim \alpha_s^4$) \cite{nnnloBSR}. We present results of the research~\cite{ACKS} into
the applicability of perturbative QCD (pQCD) with inclusion of one higher-twist (HT) contribution $\sim 1/Q^2$ dictated by OPE. At low momenta $Q^2 < Q_0^2$ ($\approx 0.4$-$0.6 \ {\rm GeV}^2$), we use a $\chi$PT-motivated ansatz~
\cite{data2B}\footnote{
\textcolor{black}{Another low $Q^2$ expression based on the light-front holographic (LFH) coupling~\cite{LFH} was also
successfully used in~\cite{ACKS}; however, discussion of the results when this particular form was used 
is beyond the scope of this short note.}}
whose first term ($\sim Q^2$) is fixed by the Gerasimov--Drell--Hearn sum rule~\cite{GDHsr}.

For the evaluation of the LT contribution we consider, in addition to pQCD, analytic frameworks of QCD (anQCD) which are a useful  
tool to evaluate physical quantities at low-momentum transfer. In these anQCD frameworks the running coupling
has no spurious (Landau) singularities, unlike pQCD
in the usual $\MSbar$-like scheme. PQCD coupling $a(Q^2)_{\rm pt} \equiv \alpha_s(Q^2)/\pi$
possesses Landau singularities at small momenta $|Q^2| \lesssim 1 \ {\rm GeV}^2$, while the general principles of quantum field theories (QFT) require that the spacelike QCD observables ${\cal D}(Q^2)$, such as current correlators and structure functions, be holomorphic (analytic) functions of $Q^2$ throughout an entire generalized spacelike region $Q^2 \in \mathbb{C} \backslash (-\infty, 0]$.
The usual pQCD couplings $a_{\rm pt}(Q^2)$ do not reflect these properties, hence the LT part of ${\cal D}(Q^2)$ evaluated in terms of $a_{\rm pt}(Q^2)$ (as a truncated perturbation series) does not have these properties dictated by QFT.
However, in anQCD we have $a_{\rm pt}(Q^2) \mapsto \A(Q^2)$, where
$\A(Q^2)$ is the anQCD coupling holomorphic in 
$Q^2 \in \mathbb{C} \backslash (-\infty, 0]$. 
Therefore, the evaluation of the LT contribution 
${\cal D}_{\rm eval.}(Q^2) \mapsto  {\cal F}(\A(Q^2))$
has the correct holomorphic properties (the same is true for the HT contribution). 

We consider three different anQCD frameworks.  
One is the Analytic Perturbation Theory (APT) of
Shirkov {\it et al.\/}~\cite{ShS}, in which the discontinuity function $\rho_1^{\rm (pt)}(\sigma) \equiv {\rm Im}\, a_{\rm pt}(Q^2=-\sigma - i \epsilon)$
of the underlying QCD is kept unchanged.
The coupling $\A^{\rm (APT)}(Q^2)$ [the APT analog of $a_{\rm pt}(Q^2)$]  
for $Q^2 \in \mathbb{C} \backslash (-\infty, 0]$ is then obtained by
using a dispersion relation involving 
${\rm Im}\, a_{\rm pt}(Q^2=-\sigma - i \epsilon)$ in the entire interval $0 \leq \sigma < \infty$.
The APT-analogs of the 
powers $a_{\rm pt}(Q^2)^{\nu}$, $\A_{\nu}^{\rm (APT)}(Q^2)$,
were obtained in  the works ~\cite{BMS05,BMS06}; this is known as Fractional APT (FAPT).

Other considered analytic frameworks are the Two-delta analytic QCD (2$\delta$anQCD \cite{2danQCD}) and the (lattice-motivated) Three-delta analytic QCD (3$\delta$anQCD \cite{3l3danQCD,4l3danQCD}). They are less closely (than APT) based on the underlying pQCD coupling $a_{\rm pt}(Q^2)$: the equality ${\rm Im}\, \A(-\sigma - i \epsilon) = \rho_1^{\rm (pt)}(\sigma)$ is enforced only for sufficiently large $\sigma \geq M_0^2$  (where $M_0 \sim 1$ GeV is a ``pQCD-onset'' scale). For $0 < \sigma < M_0^2$, the otherwise unknown discontinuity function $\rho_1(\sigma) \equiv {\rm Im}\, \A(Q^2=-\sigma - i \epsilon)$ is parametrized by two or three delta functions, respectively. Such a parametrization is partly motivated by the Pad\'e approximant approach to the coupling $\A(Q^2)$.  
The renormalization schemes of the underlying pQCD coupling in 2$\delta$anQCD are constrained by the requirement that acceptable values of $M_0 \sim 1$ GeV and $\A(0) \sim 1$ are obtained \cite{2danQCD,anOPE,mathprg}.
In 3$\delta$anQCD, the lattice calculations provide two conditions for the coupling at very low $Q^2 < 1 \ {\rm GeV}^2$, which give two additional constraints on the delta functions. In both 2$\delta$anQCD and 3$\delta$anQCD the coupling $\A(Q^2)$ practically agrees with the underlying pQCD coupling $a_{\rm pt}(Q^2)$ at $|Q^2| \gg \Lambda^2_{\rm QCD}$, namely: $[\A(Q^2)-a_{\rm pt}(Q^2)] \sim (\Lambda^2_{\rm QCD}/Q^2)^5$. This is not the case in (F)APT where  $[\A(Q^2)-a_{\rm pt}(Q^2)] \sim (\Lambda^2_{\rm QCD}/Q^2)^1$.
The construction of the analytic analogs $\A_{n}(Q^2)$ of powers $a_{\rm pt}(Q^2)^n$, for general anQCD, was formulated for integer $n$ in~\cite{Cvetic:2006gc}, and for general (noninteger) $n$ in~\cite{GCAK}.

At $Q^2=0$, $\A(Q^2)$ is finite and positive in (F)APT and in 2$\delta$anQCD, $\A(0) \approx 0.44$ and $0.66$, respectively ($N_f=3$). Lattice results \cite{LattcoupNf0} suggest the condition $\A(0)=0$ which is enforced in 3$\delta$anQCD. Interestingly, the holomorphic coupling of Refs.~\cite{mes2} also has $\A(0)=0$.

\section{Bjorken sum rule}
\label{sec:BSR}


BSR  $\Gamma_1^{p-n}$ is the nonsinglet combination given by the difference between the proton and neutron polarized structure functions and integrated over the entire Bjorken-$x$ interval 
\be
\Gamma_1^{p-n}(Q^2)=\int_0^1 dx \left[g_1^p(x,Q^2)-g_1^n(x,Q^2) \right]\ .
\label{BSRdef}
\ee 
This is the first moment of the nonsinglet contribution to the polarized structure functions.

BSR can be written in terms of a sum of two terms,
one coming from pQCD as an expansion of the running coupling $a_{\rm pt}(Q^2)=\alpha_s(Q^2)/\pi$ and the other is the twist-4 contribution dictated by the OPE \cite{BjorkenSR}
\be
\label{BSR}
\Gamma_1^{p-n}(Q^2)=\frac{g_A}{6}E_{\rm {NS}}(Q^2)+ \frac{\mu_{4}^{p-n}(Q^2)}{Q^{2}}\ .
\ee
In the limit $Q^2\to\infty$ we have $\Gamma_1^{p-n}(\infty)=g_A/6$, where $g_A$ is the nucleon axial charge, $g_A=1.2723\pm0.0023$ \cite{PDG2016}. Higher-twist contributions are neglected here.

If we use the OPE formalism, the elastic contribution (at $x=1$) to BSR (\ref{BSRdef}) should in principle be included. It is convenient to exclude the elastic contribution, since the $Q^2$-dependence of the nonsinglet inelastic BSR at low $Q^2$ is constrained by the Gerasimov--Drell--Hearn (GDH) sum rule~\cite{GDHsr}, as was highlighted in~\cite{PSTSK10}. Further, at high $Q^2$ the elastic contribution is not noticable \cite{PSTSK10}. Hence we investigate the behavior of the pure inelastic contribution as a continuation to low-energy regime \cite{GDHlow}.

The twist-2 contribution $E_{\rm {NS}}(Q^2)$
in (\ref{BSR}) is known up to N$^3$LO~\cite{nnnloBSR}:
\be
E_{\rm NS}(Q^2)=1+e_1^{\rm NS} a_{\rm pt}(Q^2) + 
e_2^{\rm NS} a_{\rm pt}(Q^2)^2 + 
e_3^{\rm NS} a_{\rm pt}(Q^2)^3 + 
e_4^{\rm NS} a_{\rm pt}(Q^2)^4\ .
\label{Ens}
\ee  
We will evaluate this in analytic QCD frameworks, where the perturbation 
series (\ref{Ens}) must be expressed as a nonpower series via the transformation 
$a(Q^2)^n\mapsto\A_n^{(j)}(Q^2)$, 
where index $j$ indicates the analytic QCD framework ($j=$APT, $2\delta$anQCD, and $3\delta$anQCD). For the numerical evaluation
of $\A_n^{(j)}(Q^2)$'s, various programs~\cite{2danQCD,3l3danQCD,4l3danQCD,mathprg} written in Mathematica are used.\footnote{
The programs can be downloaded from the web page 
http://gcvetic.usm.cl}

We will use $N_f=3$ and various renormalization scales (RScl) $\mu^2 \not= Q^2$ in the evaluation of the quantity $E_{{\rm NS},j}(Q^2)$. In such a case, the dependence on the RScl parameter $C \equiv \ln(\mu^2/Q^2)$ enters the coefficients $e_j^{\rm NS} \mapsto e_j^{\rm NS}(C)$ and the couplings $\A_n^{(j)}(Q^2 \exp(C))$. 

The last term in BSR (\ref{BSR})  has known evolution \cite{ShuVa} in pQCD and, consequently, in general analytic versions
\be
\mu_{4}^{p-n}(Q^2)=\mu_{4}^{p-n}(Q_{\rm in}^2)\left(\frac{a_{\rm pt}(Q^2)}
{a_{\rm pt}(Q_{\rm in}^2)}\right)^{\gamma_0/8\beta_0}\ ,~~
\mu_{4,j}^{p-n}(Q^2)=\mu_{4,j}^{p-n}(Q_{\rm in}^2) \frac{\A_{\gamma_0/8\beta_0}^{(j)}(Q^2)}
{\A_{\gamma_0/8\beta_0}^{(j)}(Q_{\rm in}^2)}\ .
\label{HTQ2}
\ee
These contributions 
are important in the low-energy regime $Q^2\sim 1 \ {\rm GeV}^2$. At very low $Q^2 < 1 \ {\rm GeV}^2$, they
grow quickly and the OPE series diverges. This is a general problem in OPE. However, we can address this problem by replacing the OPE expression (\ref{BSR}) at very low $Q^2 < Q_0^2$ ($\approx 0.5 \ {\rm GeV}^2$) with a $\chi$PT-motivated expression~\cite{data2B}\footnote{There are also other methods to address the low-$Q^2$ regime, e.g., an extension from the GDH sum rule made via a QCD-improved model~\cite{GDHlow}, or with a resummation of perturbative series in~\cite{Kotikov12}, or with the LFH QCD-motivated ansatz \cite{LFH}.}:
\be
\Gamma_1^{p-n}(Q^2) = \frac{(\kappa_n^2-\kappa_p^2)}{8 M_p^2} Q^2 + A \; (Q^2)^2 + B \; (Q^2)^3
\quad (Q^2 \lesssim 0.5 \ {\rm GeV}^2).
\label{BSRlow}
\ee
Here, $\kappa_X$ is the anomalous moment of the nucleon $X$ ($\kappa_p=1.793$, $\kappa_n=1.916$), $A$ and $B$ are fit parameters. The first term ($\sim Q^2$) originates from the Gerasimov--Drell--Hearn sum rule~\cite{GDHsr}. At higher $Q^2$ 
$(Q^2 \gtrsim 0.5 \ {\rm GeV}^2)$,
we use OPE (\ref{BSR}).

\section{Numerical Results}
\label{sec:num}


Here we first test only the OPE approach (\ref{BSR}) with pQCD $\MSbar$
LT term. We search at each order for a minimum scale $Q^2=Q^2_{\rm min}$ where $\chi^2$ is minimal, while the other fit parameter is $\mu_4^{p-n}$. 
In order to fix the $\MSbar$ QCD scale ${\overline \Lambda}$, we perform the standard extraction, i.e., 
${\overline \Lambda}_{\rm (N_f=3)}$ value is obtained from a reference value, $a(Q^2=M_Z^2)=0.1185/\pi$ \cite{PDG2014}. 
The RGE evolution of $a_{\rm pt}(Q^2)$  down from $Q^2=M_Z^2$ ($N_f=5$) to low $Q^2$ (where $N_f=3$) is carried out with four-loop $\MSbar$ beta function
and with the corresponding three-loop quark threshold conditions \cite{CKS}
(at thresholds $\kappa {\overline m}_q( {\overline m}_q)$ with $\kappa=1$).
\begin{table}
\caption{\footnotesize The HT coefficient $\mu_4^{p-n}(Q^2_{\rm in})$  ($Q^2_{\rm in}=1 \ {\rm GeV}^2$) and $Q^2_{\rm min}$, both in ${\rm GeV}^2$, for pQCD extracted from data at $Q^2 \leq 3 \ {\rm GeV}^2$, in various perturbative QCD orders (up to N$^3$LO).  }
\label{tabHTpQ}
\centering
\begin{tabular}{r|cccc}
\hline
pQCD & LO & NLO & N$^2$LO & N$^3$LO
\\
\hline
$\mu_{4,pQCD}^{p-n}(1.)$ & -0.059$\pm$0.002 & -0.037$\pm$0.002 & -0.031$\pm$0.002 & -0.008$\pm$0.002 \\
$Q^2_{\rm min}$ & 0.660 & 0.660 & 0.844 & 0.844 \\
\hline
\end{tabular}
\end{table}

In Table~\ref{tabHTpQ} we present the obtained values of $\mu_{4,pQCD}^{p-n}$ (only the statistical errors were considered), and $Q^2_{\rm min}$, to various orders in the perturbation expansion (\ref{Ens}). 
As noted in previous works \cite{PSTSK10,Kataev:1997nc,Parente:1994bf}, a duality between HT contribution and the order of a perturbation series appears: when we use higher order in pQCD, the HT contribution becomes smaller in its absolute value. But this apparent property is unstable, because the $\mu_{4,pQCD}^{p-n}$ coefficient is very sensitive to $\Lambda^{(\rm pQCD)}$ parameter at N$^3$LO \cite{PSTSK10}.
The extracted values are consistent with those obtained  at LO in \cite{Ross:1993gb},  and at NLO in \cite{Sidorov:2006vu}. 

\begin{figure}[htb] 
\centering\includegraphics[width=122mm]{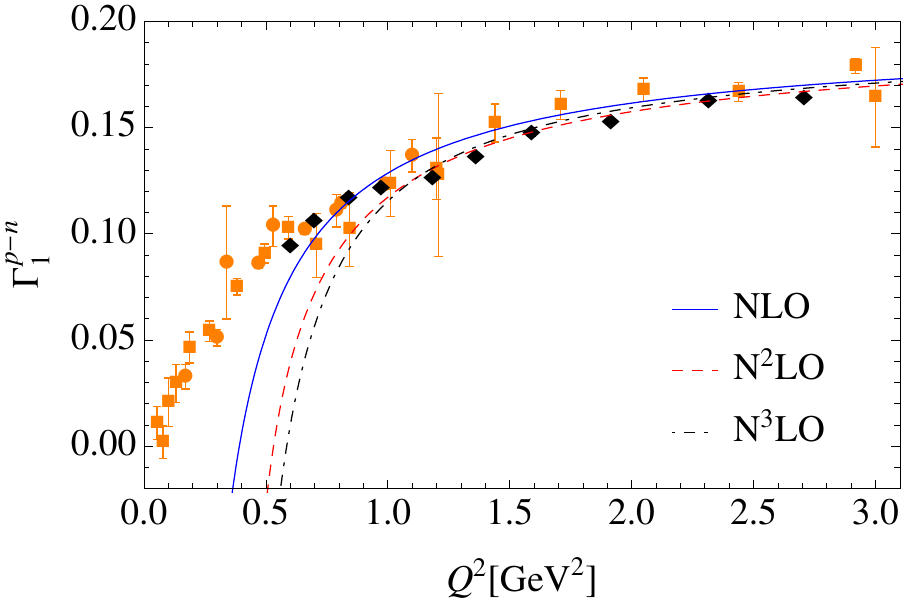} 
\vspace{0.0cm} 
\caption{The described fits of data \cite{data2,data2B,data2N,data3} on BSR $\Gamma_1^{p-n}$ as a function of $Q^2$, to various orders of perturbation series (\ref{BSR}). The newer data~\cite{data2N} have small statistical errors and are in black, and the older data \cite{data2,data2B,data3} are in light grey (orange online).}
\label{figpQCD} 
\end{figure}

In Fig.~\ref{figpQCD} we show the pQCD fit of BSR function $\Gamma_1^{p-n}$ at NLO, N$^2$LO and N$^3$LO. Increasing the perturbation order, the range of applicability of pQCD becomes smaller, covering fewer points of data in the low-$Q^2$ region.

\begin{table}
  \caption{\footnotesize The fit parameter values obtained with various approaches to BSR.}
\label{TabResNf3}
\centering
\begin{tabular}{r|llllll}
\hline
Approach ($j$) & $\mu_{4,j}^{p-n}(1.)$ & $C$ & $Q_0^2$ & $A$ & $B$ & $\chi^2$
\\
\hline
$\MSbar$ pQCD     & -0.0344 & 1.801 & 0.646 & 0.658 & -0.840 & 24.44
\\
(F)APT            & -0.0498 & 1.019 & 0.633 & 0.658 & -0.840 & 13.53
\\
2$\delta$anQCD    & -0.0238 & -0.859 & 0.500 & 0.831 & -1.269 & 5.49
\\
(3l)3$\delta$anQCD & -0.0105 & 0.795 & 0.467 & 0.752 & -1.065 & 4.97
\\
(4l)3$\delta$anQCD & -0.0187 & 1.017 & 0.431 & 0.842 & -1.342 & 4.95\\
\hline
\end{tabular}
\end{table}

Now we employ analytic (holomorphic) QCD approaches in the fits.
In Table \ref{TabResNf3} we show, for five different evaluations of the LT contribution $E_{\rm {NS}}(Q^2)$, the resulting values of the fit parameters: HT coefficient $\mu_4^{p-n}(Q^2_{\rm in})$; RScl parameter $C \equiv \ln(\mu^2/Q^2)$ of the LT contribution; matching point scale $Q_0^2$; parameter $A$ in the $\chi$PT-motivated expression (\ref{BSRlow}).
The values of the $B$ parameter of the $\chi$PT-motivated expression were extracted by the matching condition at $Q^2=Q_0^2$. The last column gives the values of $\chi^2$ for the obtained curves.
The resulting values of $A$ are approximately consistent with the value $A=0.74$ obtained in $\chi$PT calculations in~\cite{Jietal} but not with the value  $A=2.4$ obtained in~\cite{Beretal}. 

 \begin{figure}[htb] 
\centering\includegraphics[width=140mm,height=90mm]{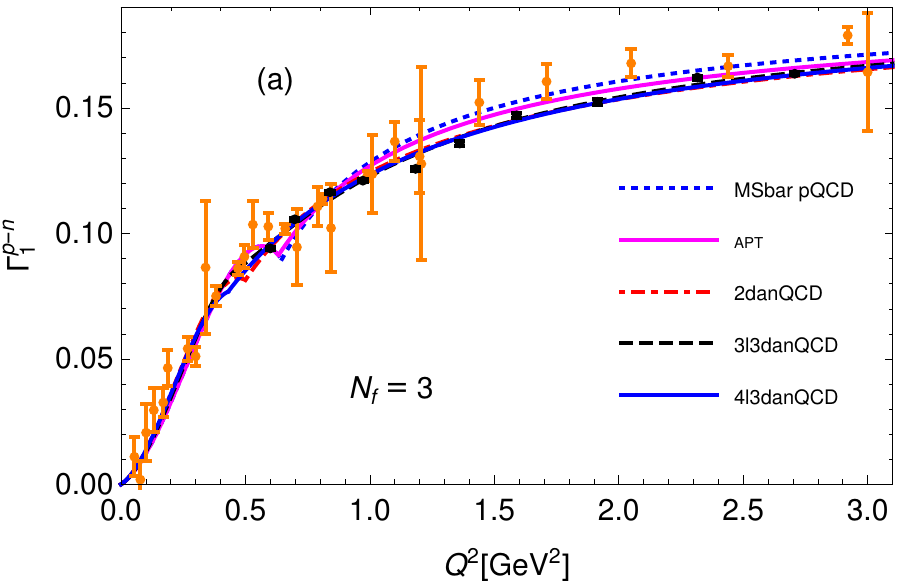}
\vspace{-0.4cm}
\caption{
Fits of data \cite{data2,data2B,data2N,data3} on BSR $\Gamma_1^{p-n}(Q^2)$ 
as a function of $Q^2$, using at $Q^2 \geq Q_0^2$ (four-loop) $\MSbar$ pQCD and analytic QCD frameworks.
} 
\label{FigFitNf3}
\end{figure}
The obtained curves are shown in Fig.~\ref{FigFitNf3}. As mentioned, these curves consist of two curves ``glued together'' at a matching point $Q_0^2$: the OPE curve (\ref{BSR}) for $Q^2 \geq Q_0^2$ and the $\chi$PT-motivated curve (\ref{BSRlow}) for $Q^2 \leq Q_0^2$. 

These curves indicate that the pQCD $\MSbar$ approach and, to a lesser extent, the (F)APT approach, give a visible slope discontinuity (kink) at the matching point $Q^2=Q_0^2$ between the OPE and the $\chi$PT-motivated expression, i.e., they cannot bridge well the gap between the high and low-$Q^2$ regimes. However, 2$\delta$anQCD and 3$\delta$anQCD are able to bridge this gap without a visible kink, cf.~Fig.~\ref{FigFitNf3}.

\section{Conclusions}
\label{sec:concl}

In~\cite{ACKS} we investigated the Bjorken polarized sum rule (BSR) $\Gamma_1^{p-n}(Q^2)$ (with the elastic contribution excluded) as a function of squared momentum transfer $Q^2$, at $Q^2 \leq 3 \ {\rm GeV}^2$ in various QCD approaches, and compared it with the available experimental data. For $Q^2 \geq Q_0^2$ ($\approx 0.5 \ {\rm GeV}^2$) we used the theoretical expressions for the leading-twist (LT) contribution plus one higher-twist (HT) term $\mu_4^{p-n}/Q^2$, Eqs.~(\ref{Ens}) and (\ref{BSR}). For $Q^2 \leq Q_0^2$, the $\chi$PT-motivated expression 
(\ref{BSRlow}) was used. The fit parameters were the HT coefficient $\mu_4^{p-n}(Q^2_{\rm in})$ (at $Q^2_{\rm in}=1 \ {\rm GeV}^2$), the renormalization scale (RScl) parameter $C \equiv \ln(\mu^2/Q^2)$ in the LT contribution, the transition scale $Q_0^2$, and the free parameter $A$ in the $\chi$PT-motivated expression (\ref{BSRlow}). For the evaluation of the LT contribution at $Q^2 \geq Q_0^2$ we used the usual $\MSbar$ pQCD, and various QCD versions with holomorphic infrared-safe coupling $\A(Q^2)$: (F)APT \cite{ShS}; 2$\delta$anQCD \cite{2danQCD,anOPE,mathprg}; and a lattice-motivated 3$\delta$anQCD coupling in the three- and four-loop lattice MiniMOM scheme: 3l3$\delta$anQCD \cite{3l3danQCD} and 4l3$\delta$anQCD \cite{4l3danQCD}. It turned out that the latter three holomorphic (analytic) QCD versions give the best fit results and the lowest values of $Q_0^2 \approx 0.4$-$0.5 \ {\rm GeV}^2$ and $\chi^2 \approx 5.0$-$5.5$. (F)APT requires a higher value $Q_0^2 \approx 0.63 \ {\rm GeV}^2$ and gives $\chi^2 \approx 13.5$. The $\MSbar$ pQCD gives the worst results, $\chi^2 \approx 24.$; the principal reason for this lies in the fact that the $\MSbar$ pQCD coupling $a_{\rm pt}(Q^2)$ has Landau singularities at positive $Q^2 \leq 0.37 \ {\rm GeV}^2$, and this makes the evaluation of low-$Q^2$ BSR unreliable.

The presented evaluation of low-$Q^2$ BSR shows that it is imperative to use QCD frameworks whose couplings have no Landau singularities in this region.

\ack
The work of C.A. was supported by FONDECYT Postdoctoral Grant No.~3170116. The work of A.V.K. and B.G.S. was supported in part by the RFBR 
Grant No. 16-02-00790-a. \\

\end{document}